\documentclass[twocolumn,
amsmath,amssymb,prd,preprintnumbers,superscriptaddress,showpacs]{revtex4-1}

\usepackage{graphicx}
\usepackage{latexsym}
\usepackage{bm}
\usepackage{slashed}

\def\half{{\textstyle\frac12}}

\def\lsim{\mathrel{\rlap{\lower3pt\hbox{$\sim$}}
    \raise2pt\hbox{$<$}}}
\def\gsim{\mathrel{\rlap{\lower3pt\hbox{$\sim$}}
    \raise2pt\hbox{$>$}}}
\def\sqr#1#2{{\vcenter{\vbox{\hrule height.#2pt
         \hbox{\vrule width.#2pt height#1pt \kern#1pt
         \vrule width.#2pt}
         \hrule height.#2pt}}}}

\def\prt{\partial}
\def\lrpartial{\raise 1pt\hbox{$\stackrel\leftrightarrow\partial$}}

\newcommand{\beq}[1]{\begin{equation}\label{#1}}
\newcommand{\eeq}{\end{equation}}
\newcommand{\bea}[1]{\begin{eqnarray}\label{#1}}
\newcommand{\eea}{\end{eqnarray}}

\newcommand{\rf}[1]{(\ref{#1})}

\def\etal{{\it et al.}}

\begin{document}

\title{Massive photons and Lorentz violation}

\author{Mauro Cambiaso}
\affiliation{Universidad Andres Bello, Departamento de Ciencias Fisicas,
Facultad de Ciencias Exactas, Avenida Republica 220, Santiago, Chile}

\author{Ralf Lehnert}
\affiliation{Indiana University Center for Spacetime Symmetries, 
Bloomington, IN 47405, USA}

\author{Robertus Potting}
\affiliation{CENTRA, Departamento de F\'\i sica, Universidade do Algarve, 8005-139 Faro, Portugal}

\date{January 13, 2012}

\begin{abstract} 
All quadratic translation- and gauge-invariant photon operators
for Lorentz breakdown
are included into the Stueckelberg Lagrangian 
for massive photons in a generalized $R_\xi$ gauge.
The corresponding dispersion relation and tree-level propagator are determined exactly, 
and some leading-order results are derived. 
The question of how to include such Lorentz-violating effects 
into a perturbative quantum-field expansion is addressed.
Applications of these results within Lorentz-breaking quantum field theories
include the regularization of infrared divergences
as well as the free propagation of massive vector bosons.
\end{abstract}

\pacs{11.30.Cp, 11.30.Er, 11.30.Qc, 12.20.-m, 12.60.Cn}

\maketitle

\section{introduction}
\label{intro}

Recent years have witnessed a growing interest
in precision tests of Lorentz and CPT invariance.
This interest can be partly attributed to 
the availability of new observational data 
and the development of ultra-sensitive experimental techniques~\cite{ExpRev}.
Moreover, 
minute departures from Lorentz and CPT symmetry can be accommodated in
various theoretical ideas beyond established physics~\cite{motivation}
providing a phenomenological opportunity to search for novel effects 
possibly arising at the Planck scale. 

At presently attainable energies, 
Lorentz- and CPT-violating effects are expected to be governed by
effective field theory~\cite{EFT}.
The general framework based on this premise
is known as the Standard-Model Extension (SME)~\cite{SMEpapers,caus,PhotonSME}.
This framework contains 
the usual Standard Model of particle physics and general relativity 
as limiting cases 
and therefore permits the identification and analysis
of essentially all currently feasible Lorentz and CPT tests.
To date, 
the SME has been employed for phenomenological studies involving
cosmic radiation~\cite{UHECR},
meson factories~\cite{mesons} and other particle colliders~\cite{collider},
resonance cavities~\cite{cavities},
neutrinos~\cite{neutrinos}, 
precision spectroscopy~\cite{spectroscopy},
and gravity~\cite{GravExp}.

The SME has also served as the basis for
various theoretical investigations of Lorentz and CPT symmetry.
These investigations have have shed light on 
the SME's mathematical structure~\cite{math},
spontaneous Lorentz breakdown and Nambu--Goldstone modes~\cite{SSB},
classical limits of Lorentz- and CPT-violating physics~\cite{classical},
quantum corrections and renormalizability~\cite{loops,AACGS},
features of non-renormalizable contributions~\cite{nonren},
etc.
At the same time, 
these analyses have solidified 
various aspects of the SME's theoretical foundation.

One topic that has remained comparatively unexplored 
concerns Lorentz and CPT violations in massive vector particles in the SME:
published work~\cite{EWpaper} has been focused on
phenomenological analyses confined to the CPT-even $Z^0$ and $W^\pm$ sectors 
of the minimal SME.
However, 
even for this small region in the full SME's parameter space 
basic theoretical results,
such as expressions for the dispersion relation and propagator,  
are currently still lacking. 
The present investigation reports on progress towards filling this gap.

Massive vector particles with Lorentz and CPT breaking 
are not only of interest for phenomenological studies of
the heavy gauge bosons $Z^0$ and $W^\pm$.
They have previously served a valuable tool for 
investigations of 
the mass-dimension three Lorentz- and CPT-violating Maxwell--Chern--Simons term~\cite{AACGS}.
More importantly, 
massive vector fields play a key role for 
theoretical studies involving the photon 
because they provide a popular method for 
regularizing infrared divergences in perturbative quantum-field calculations.
In fact, 
this latter application represents the primary focus 
of this study.
But we anticipate that
our results for the expression of the dispersion relation and the propagator 
are equally valid for 
the quadratic part of the full SME's $Z^0$ and $W^\pm$ sectors.

The outline of the present work is as follows.
Section~\ref{fullSME} 
provides the basic ideas behind the construction of the model 
we are studying.
A few remarks on the structure of the resulting 
modified Maxwell--Stueckelberg equations 
are contained in Sec.~\ref{results}.
Section~\ref{SolFeat} determines 
the exact dispersion relation and propagator for our model.
Section~\ref{rules} discusses some leading-order results 
that are expected to be useful for practical calculations, 
and Sec.~\ref{summary} gives a brief summary of our results. 
Some supplementary material is collected in various appendices.


\section{Model Basics}
\label{fullSME}

Our primary goal is to introduce a photon mass for 
regularizing infrared divergences in perturbative quantum-field calculations 
in a general $R_\xi$ gauge.
This requires 
a smooth behavior of the internal-symmetry structure 
in the massless limit. 
Thus, 
the usual Proca term by itself is insufficient,
and the mass needs to be introduced via, 
e.g., 
the Stueckelberg method~\cite{Stueckelberg}. 
The original version of this method is appropriate for U(1) gauge theories.
The method needs modifications for photons embedded in the U(1)$\times$SU(2) 
gauge structure of the Standard Model~\cite{StueckelbergSM},
and it generally fails for non-Abelian vector fields. 
The Lorentz-violating generalization of the Stueckelberg method, 
which is discussed in this section and the subsequent one, 
therefore applies solely to the SME's QED limit. 
We do, 
however, 
expect the remaining part of our study, 
contained in Secs.~\ref{SolFeat} and~\ref{rules}, 
to be applicable to more general massive vector fields.

Our starting point is the usual 
free-photon Lagrangian
minimally coupled to an external conserved current $j^\mu$.
The generalization of this Lagrangian
to include arbitrary local, 
coordinate-independent,
translation- and gauge-invariant physics
with Lorentz- and CPT-symmetry breakdown 
can be cast into the following form~\cite{PhotonSME}:
\bea{lagrangian}
{}\hspace{-5mm}{\cal L}_{\gamma} \!& = &\!
-\tfrac{1}{4}F^2
-A\cdot j\nonumber\\
&&{}-\tfrac{1}{4}F_{\kappa\lambda}(\hat{k}_F)^{\kappa\lambda\mu\nu}F_{\mu\nu}
+\tfrac{1}{2}\epsilon^{\kappa\lambda\mu\nu}A_\lambda(\hat{k}_{AF})_\kappa F_{\mu\nu}\,.
\eea
Here,
the field strengths and potentials are real-valued and 
obey the conventional relation
$F_{\mu\nu}=\partial_\mu A_\nu-\partial_\nu A_\mu$,
and $\epsilon^{\kappa\lambda\mu\nu}$ denotes the totally antisymmetric symbol 
with $\epsilon^{0123}=+1$.
Lorentz and CPT breakdown is controlled by 
the quantities $\hat{k}_F$ and $\hat{k}_{AF}$, 
which are given explicitly by the following expressions~\cite{PhotonSME}
\bea{smecoeff}
{}\hspace{-5mm}(\hat{k}_F)^{\kappa\lambda\mu\nu} \!& = &\!
\sum_{n=2}^{\infty}(k_F^{(2n)})^{\kappa\lambda\mu\nu\alpha_1 \ldots \alpha_{2n-4}}
\partial_{\alpha_1} \ldots \partial_{\alpha_{2n-4}}\nonumber\\
(\hat{k}_{AF})_{\kappa} \!& = &\! \sum_{n=1}^{\infty} 
(k_{AF}^{(2n+1)})_{\kappa}{}^{\alpha_1\ldots\alpha_{2n-2}}
\partial_{\alpha_1} \ldots \partial_{\alpha_{2n-2}}\,.
\eea
Each $k_F^{(d)}$ coefficient as well as each $k_{AF}^{(d)}$ coefficient 
is taken as nondynamical, 
spacetime constant, 
and totally symmetric in its $\alpha$ indices.
The superscript $(d)$ labels the mass dimension of the corresponding photon operator, 
so that the unit of the actual coefficient is $[\textrm{GeV}]^{\,4-d}$.

The next step is to add a mass-type term $\delta {\cal L}_{m}$ 
for the photon  
to the above Lagrangian~\rf{lagrangian}.
In the conventional case, 
such a contribution is restricted by Lorentz symmetry.
In the present situation,
this restriction is absent, 
and more freedom in the choice of $\delta {\cal L}_{m}$ exists.
This additional freedom partly depends on 
the type of physics to be described. 
For instance, 
one may wish to model general Lorentz violation 
for hypothetical massive photons. 
Alternatively, 
the aim may be to regularize infrared divergences 
that often arise in quantum-field calculations
involving massless photons 
governed by Lagrangian~\rf{lagrangian}.
In these two examples, 
the former may allow more additional freedom for $\delta {\cal L}_{m}$
than the latter: 
consider a situation in which the coefficients in Lagrangian~\rf{lagrangian} 
are such that a subgroup of the Lorentz group remains intact.
A regulator breaking this residual symmetry 
may be problematic, 
so that violations of the remaining invariant subgroup
may have to be excluded from $\delta {\cal L}_{m}$. 

In the present work,
the primary purpose for the introduction of a photon-mass term 
is to regularize potential infrared divergences in quantum-field contexts. 
In principle, 
the structure of $\delta {\cal L}_{m}$
can then be chosen as simple as possible. 
For example, 
the conventional Lorentz-symmetric Stueckelberg expression would likely suffice.

However, 
we introduce an additional set of Lorentz-breaking coefficients 
for reasons outside the present U(1) context:
certain aspects of the Stueckelberg model, 
such as the dispersion relation and propagator,
will turn out to be equally valid for the Lorentz-violating $Z^0$ and $W^\pm$ bosons.
These particles contain not only the equivalent of the $\hat{k}_F$ and $\hat{k}_{AF}$ coefficients, 
but also, 
e.g.,
a $(k_{\phi\phi})^{\mu\nu}$ mass-type term. 
For wider applicability in this electroweak context, 
we therefore also include a $(k_{\phi\phi})^{\mu\nu}$-type contribution 
into our Stueckelberg mass term.
With these considerations in mind,
we implement the Stueckelberg method~\cite{Stueckelberg} by
introducing a scalar field $\phi$ in the following way:
\beq{Stueckelberg}
\delta {\cal L}_{m}=
\tfrac{1}{2}(\partial_\mu \phi-m A_\mu)
\hat{\eta}^{\mu\nu}(\partial_\nu \phi-m A_\nu)\,,
\eeq
where $m$ denotes the photon mass
and
\beq{hat-eta}
\hat{\eta}^{\mu\nu}=\eta^{\mu\nu}+\hat{G}^{\mu\nu}.
\eeq
Here, 
the $\hat{G}^{\mu\nu}$ represents the full-SME generalization 
of the minimal-SME's $(k_{\phi\phi})^{\mu\nu}$ coefficient 
and is given by:
\beq{Gdef}
\hat{G}^{\mu\nu} = 
\sum_{n=2}^{\infty}(G^{(2n)})^{\mu\nu\alpha_1 \ldots \alpha_{2n-4}}
\partial_{\alpha_1} \ldots \partial_{\alpha_{2n-4}}\,,
\eeq
where each $(G^{(2n)})^{\mu\nu\alpha_1 \ldots \alpha_{2n-4}}$ 
is contracted with an even number of derivatives~\cite{fn1},
is spacetime constant, 
symmetric in $\mu$ and $\nu$,
and totally symmetric in $\alpha_1 \ldots \alpha_{2n-4}$.
This definition still contains some Lorentz-symmetric pieces 
at each mass dimension, 
which can be eliminated if necessary. 
For example, 
the Lorentz-covariant contribution $\sim\eta^{\mu\nu}$ 
contained in $(G^{(4)})^{\mu\nu}$ 
can be removed 
by taking this coefficient as traceless.

The inclusion of $\hat{G}^{\mu\nu}$
does not invalidate the Stueckelberg method.
At this point, 
we leave $\hat{G}^{\mu\nu}$ undetermined.
We only require it to be small, 
so that in the limit of vanishing Lorentz violation, 
$\hat{\eta}^{\mu\nu}$ approaches $\eta^{\mu\nu}$ 
without a change in signature and rank~\cite{fn2}.
In the present U(1) context, 
where $\delta {\cal L}_{m}$ is intended to serve as a regulator, 
specific regions in ($k_F^{(d)}$, $k_{AF}^{(d)}$) parameter space
may only be compatible with certain definite choices for $\hat{G}^{\mu\nu}$, 
such as $\hat{G}^{\mu\nu}=0$, 
as discussed above. 
For applications in the heavy-boson context, 
$\hat{G}^{\mu\nu}$ represents an arbitrary physical parameter that
can only be fixed by observation.

As in the ordinary case,
the resulting Lagrangian ${\cal L}_{m}\equiv {\cal L}_\gamma+\delta {\cal L}_{m}$ 
changes under a local gauge transformation
\beq{eq:gaugetransf}
\delta A_\mu = \partial_\mu \epsilon(x)\,,\quad \delta \phi = m \epsilon(x)
\eeq
by total-derivative terms.
However,
in the absence of topological obstructions 
and with the usual boundary conditions, 
the action
and thus the physics remain unchanged 
under the transformation~\rf{eq:gaugetransf}.
The next natural step then is to select a gauge-fixing condition $\mathcal{F}[A,\phi]$.
As usual,
a multitude of choices for $\mathcal{F}[A,\phi]$ are acceptable. 
We take 
\beq{GaugeCondition} 
\mathcal{F}[A,\phi]=\prt_\mu \hat{\eta}^{\mu \nu}A_\nu + \xi m\phi\,,
\eeq
a choice that will turn out to be convenient for our purposes.
Application of the usual Gaussian smearing procedure 
leads to the following gauge-fixing term 
to be included into the Lagrangian:
\beq{lagrangian_GF}
\mathcal{L}_\textrm{g.f.} = -{1 \over 2\xi} (\prt_\mu \hat{\eta}^{\mu \nu}A_\nu + \xi m \phi)^2\,,
\eeq
where $\xi$ is an arbitrary gauge parameter,
as usual.
The associated Faddeev--Popov determinant 
$\det\!\left(\frac{\delta\mathcal{F}}{\delta\epsilon}\right)$ in the path integral
results in the ghost term
\begin{equation} \label{lagrangian_FP}
\mathcal{L}_\textrm{F.P.} = -\bar c(\prt_\mu\hat{\eta}^{\mu\nu}\prt_\nu+\xi m^2)c\,,
\end{equation}
where $c$ and $\bar c$ are anticommuting scalars.
We mention that 
the possibility of introducing additional Lorentz violation 
into the ghost Lagrangian has been studied~\cite{AltschulFP}.
We disregard this option in what follows.

We are now in the position to present our model Lagrangian 
${\cal L}={\cal L}_\gamma+\delta {\cal L}_{m}+\mathcal{L}_\textrm{g.f.}+ \mathcal{L}_\textrm{F.P.}$
explicitly:
\begin{eqnarray}
\label{eq:totalL}
{}\hspace{-10mm}\mathcal{L}\! &=&\!-\tfrac{1}{4}F^2
-A\cdot j +\tfrac{1}{2}m^2 A_\mu \hat{\eta}^{\mu \nu} A_\nu
-{1\over2\xi}(\prt_\mu \hat{\eta}^{\mu \nu}A_\nu)^2\nonumber\\
&&\! {}-\half \phi(\prt_\mu\hat{\eta}^{\mu\nu}\prt_\nu+\xi m^2)\phi
- \bar c(\prt_\mu\hat{\eta}^{\mu\nu}\prt_\nu+\xi m^2)c\frac{{}}{{}}\nonumber\\
&&\!{}-\tfrac{1}{4}F_{\kappa\lambda}(\hat{k}_F)^{\kappa\lambda\mu\nu}F_{\mu\nu}
+\tfrac{1}{2}\epsilon^{\kappa\lambda\mu\nu}A_\lambda(\hat{k}_{AF})_\kappa F_{\mu\nu}\,.
\end{eqnarray}
It is apparent that 
the scalar $\phi$ and the ghosts $c$ and $\bar c$ are now uncoupled
and can be integrated out of the path integral 
yielding an unobservable normalization constant.
We will therefore disregard these fields in our subsequent analysis.


\section{Structure of the field equations}
\label{results}

The equations of motion for our Lagrangian~\rf{eq:totalL} read
\bea{EoM1}
\hspace{-7.5mm}\left[\eta^{\mu\alpha}\eta^{\nu\beta}\partial_\mu
+(\hat{k}_{AF})_\mu\epsilon^{\mu\nu\alpha\beta}
+(\hat{k}_F)^{\mu\nu\alpha\beta}\partial_\mu\right]F_{\alpha\beta}\!&&\!\nonumber\\
{}+\left[m^2\hat{\eta}^{\mu\nu}
+\frac{1}{\xi}\hat{\eta}^{\mu\alpha}\hat{\eta}^{\nu\beta}\partial_\alpha\partial_\beta\right]A_\mu\!&=&\!j^\nu
\,.
\eea
Owing to its underlying antisymmetric structure, 
the term involving $F_{\alpha\beta}$ in Eq.~\rf{EoM1}
has vanishing divergence.
Contraction of the field equations with $\partial_\nu$
therefore removes the $F_{\alpha\beta}$ term
and places the constraint
\beq{longitudinal}
\left(\hat{\eta}^{\mu\nu}\partial_\mu\partial_\nu+\xi m^2
\right)(\partial_\alpha\hat{\eta}^{\alpha\beta}A_\beta)=0
\eeq
on the $A_\mu$ term,
where we have used our earlier assumption 
of a conserved source $j^\nu$. 
As per definition, 
$\hat{\eta}^{\alpha\beta}$ is of rank four, 
so $(\partial_\alpha\hat{\eta}^{\alpha\beta}A_\beta)$ projects out 
one of the degrees of freedom contained in $A_\mu$. 
The $\xi$-dependent Eq.~\rf{longitudinal} shows that 
the source $j_\nu$ does not excite $(\partial_\alpha\hat{\eta}^{\alpha\beta}A_\beta)$.
This degree of freedom therefore is an auxiliary mode.

Continuing with this decomposition,
it is natural to define a component 
\beq{transverse}
A^{\rm ph}_\nu\equiv A_\nu+\frac{1}{\xi m^2}\partial_\nu(\partial_\alpha\hat{\eta}^{\alpha\beta}A_\beta)\,.
\eeq
Employing Eq.~\rf{longitudinal}, 
it is apparent that 
$\partial_\mu\hat{\eta}^{\mu\nu}A^{\rm ph}_\nu=0$ on shell. 
With 
Eq.~\rf{transverse} at hand,
we can now substitute 
the decomposition of the vector potential 
$A_\nu= A^{\rm ph}_\nu-\partial_\nu(\partial_\alpha\hat{\eta}^{\alpha\beta}A_\beta)/(\xi m^2)$ 
into the field equations~\rf{EoM1}.
Being a gradient,
the auxiliary excitation does not contribute to $F_{\alpha\beta}$. 
By virtue of its equation of motion~\rf{longitudinal},
this component also disappears from the $A_\mu$ term in Eq.~\rf{EoM1}.
The zero divergence of $A^{\rm ph}_\nu$,
on the other hand, 
implies that
it vanishes when contracted with the $\xi$ term in Eq.~\rf{EoM1}.
Our decomposition then gives
\bea{EoM2}
\hspace{-7.5mm}\left[\eta^{\mu\alpha}\eta^{\nu\beta}\partial_\mu
+(\hat{k}_{AF})_\mu\epsilon^{\mu\nu\alpha\beta}
+(\hat{k}_F)^{\mu\nu\alpha\beta}\partial_\mu\right]F_{\alpha\beta}\!&&\!\nonumber\\
{}+\left[m^2\hat{\eta}^{\mu\nu}\right]A^{\rm ph}_\mu\!&=&\!j^\nu
\eea
for the field equations.
Note that the auxiliary component has disappeared entirely
($F_{\alpha\beta}$ only involves $A^{\rm ph}_\nu$)
and that the source $j^{\nu}$ excites the physical
degrees of freedom in a $\xi$-independent way.


\section{Features of the general solution}
\label{SolFeat}

In this section, 
we study general properties of the solutions
of the equation of motion~\rf{EoM1}. 
This equation holds exactly for our Lorentz-violating Stueckelberg photons
at the classical level. 
But the model also exhibits numerous similarities to 
heavy gauge bosons with Lorentz violation.
For example, 
the $Z^0$ and $W^\pm$ sectors of the SME 
also contain operators of the type $k_F$, $k_{AF}$, and $G$;
examples of these are $k_W$, $k_2$, and $k_{\phi\phi}$, 
respectively.
Although the linear equation~\rf{EoM1} 
cannot hold exactly for non-Abelian gauge bosons, 
the types of operators quadratic in the fields do agree 
with those for the photon. 
One can therefore anticipate that 
Eq.~\rf{EoM1} does govern most aspects of the tree-level free behavior 
of $Z^0$ and $W^\pm$ within the SME.
In particular, 
the dispersion relation and propagator derived below
are expected to hold not only for our modified Stueckelberg photons, 
but also for the heavy SME gauge bosons 
at tree level.

We begin with the plane-wave dispersion relation. 
To this end,
we Fourier transform Eq.~\rf{EoM1}, 
which yields
\begin{multline}
\label{EoM3}
{}\hspace{-4mm}\bigg[
p^2\eta^{\mu\nu}-p^\mu p^\nu
-m^2\hat{\eta}^{\mu\nu}
+\frac{1}{\xi}\hat{\eta}^{\mu\alpha}\hat{\eta}^{\nu\beta}p_\alpha p_\beta
\\
+2(\hat{k}_{F})^{\alpha\mu\beta\nu}p_\alpha p_\beta
-2i(\hat{k}_{AF})_\alpha\epsilon^{\alpha\beta\mu\nu}p_\beta
\bigg]\bar{A}_\nu=-\bar{\jmath}^{\,\mu}
\end{multline}
for the equations of motion
in $p^\mu$-momentum space.
Here, a bar denotes the Fourier transform,
and it is understood that the replacement $\partial\to -ip$
has been implemented in 
the Lorentz-violating quantities $\hat{k}_{F}$, $\hat{k}_{AF}$, and $\hat{G}$.
Let us briefly pause at this point 
to introduce a more concise notation that 
will enable us present many of our subsequent results 
in a more compact form.
We define
\beq{defs}
\hat{K}^{\mu\nu} \equiv (\hat{k}_{F})^{\alpha\mu\beta\nu}p_\alpha p_\beta\,,\quad
\hat{\cal E}^{\mu\nu}\equiv(\hat{k}_{AF})_\alpha\epsilon^{\alpha\beta\mu\nu}p_\beta\,,
\eeq
because in the dispersion relation and the propagator,
$\hat{k}_{F}$ and $\hat{k}_{AF}$ will always appear in this form.
Moreover, 
when convenient 
we abbreviate the contraction of a symmetric tensor with a 4-vector by 
placing the vector as a super- or subscript on the tensor, 
e.g.,
$\hat{G}^\mu_p\equiv\hat{G}^{\mu\nu} \,p_\nu$ or 
$(\hat{G}^2)^p_p\equiv\hat{G}^\mu_\alpha\hat{G}^\alpha_\nu\,p^\mu p^\nu$,
etc.
We now rewrite the equations of motion~\rf{EoM3} 
simply as $S(p)^\mu{}_\nu\, \bar{A}^\nu(p)=-\bar{\jmath}^{\,\mu}(p)$,
where 
\bea{Soperator}
S^{\mu\nu} & \equiv &
(p^2-m^2)\eta^{\mu\nu}-\left(1-\frac{1}{\xi}\right)p^\mu p^\nu
-2i\hat{\cal E}^{\mu\nu}+2\hat{K}^{\mu\nu}\nonumber\\
&&{}
-m^2\hat{G}^{\mu\nu}
+\frac{1}{\xi}\big(\hat{G}^\mu_p\, p^{\nu}+\hat{G}^\nu_p\, p^{\mu}+\hat{G}^\mu_p\hat{G}^\nu_p\big)
\eea
is the expression of the modified Stueckelberg operator 
in our new notation.

The plane-wave dispersion relation 
governs source-free motion $\bar{\jmath}^{\,\mu}=0$
and can therefore be stated as the usual requirement that 
the determinant of $S(p)^\mu{}_\nu$ vanishes.
With the results derived in Appendix~\ref{matrix}, 
this translates into the equation
\beq{DRshort}
[S]^4
-6[S]^2[S^2]
+3[S^2]^2
+8[S][S^3]
-6[S^4]=0\,.
\eeq
Here, 
$S^n$ is the $n$th matrix power of $S(p)^\mu{}_\nu$, 
and the square brackets denote the matrix trace.

The various trace expressions in Eq.~\rf{DRshort} 
can be cast into a factorized form:
\beq{fullDR}
\xi\det(S)=(\hat{\eta}^{\mu\nu}p_\mu p_\nu-\xi m^2)Q(p)=0\,.
\eeq
The $(\hat{\eta}^{\mu\nu}p_\mu p_\nu-\xi m^2)$ piece
is associated with the auxiliary mode 
described by Eq.~\rf{longitudinal}.
The $\xi$ independent factor $Q(p)$ 
governs the three physical degrees of freedom.
Both factors in the dispersion relation~\rf{fullDR} can also contain 
unphysical Ostrogradski-type degrees of freedom~\cite{Ostrogradski},
which are introduced because our Lagrangian contains higher derivatives~\cite{PhotonSME}.
In the presumed underlying theory, 
for which the effective field theory~\rf{eq:totalL} represents the low-energy limit,
these modes must be absent.
Consequently, 
they should also be eliminated from our low-energy model~\rf{eq:totalL}.

An explicit calculation shows that
\beq{physDR}
Q=(p^2-m^2)^3+
r_2(p^2-m^2)^2+
r_1(p^2-m^2)+
r_0\,,
\eeq
where the coefficients $r_j=r_j(p)$ are momentum-dependent coordinate scalars
determined by traces of combinations 
of the various Lorentz-violating tensor expressions appearing in Eq.~\rf{EoM3}.
They vanish in the limit $\hat{k}_{F},\hat{k}_{AF},\hat{G}\to0$.
The explicit expressions for the $r_j$, 
which can be found in Appendix~\ref{DR},
are not particularly transparent.
Note that in general the physical dispersion relation~\rf{physDR} 
does {\em not} represent 
a true cubic equation in the variable $(p^2-m^2)$
because the $r_j(p)$ are momentum dependent.

The dispersion relation~\rf{fullDR} 
restricts the set of all possible Fourier momenta $p^{\mu}\equiv(\omega,\vec{p})$
to those associated with plane-wave solutions of the free model. 
Since in general $\hat{k}_{F}$,
$\hat{k}_{AF}$, 
and $\hat{G}$
contain high powers of $p^{\mu}$,
there can be a corresponding multitude of plane-wave frequencies $\omega(\vec{p})$ 
for any given wave 3-vector $\vec{p}$, 
a fact reflected in the momentum dependence of the $r_j(p)$.
We remind the reader that 
most of these are artifacts of our effective-Lagrangian approach
and must be eliminated.
Only those wave momenta that 
represent perturbations of the usual Lorentz-symmetric solutions 
should be interpreted as physical. 
In any case, 
a determination of the exact roots of the general dispersion relation~\rf{fullDR} 
appears to be unfeasible. 
However, 
an exact discrete symmetry of the plane-wave solutions 
is discussed in Appendix~\ref{symmetries}, 
the massless limit is studied in Appendix~\ref{massless},
and some leading-order results are presented in Sec.~\ref{rules}.

In the more general case of non-vanishing sources,
the construction of solutions can be achieved with propagator functions. 
Paralleling the ordinary Lorentz-symmetric case,
we implicitly define the $p^\mu$-momentum space propagator $P(p)^\mu{}_\nu$
via $P(p)^\mu{}_\nu\, S(p)^\nu{}_\lambda\equiv-i\eta^\mu_\lambda$, 
where we have employed the usual quantum-field convention by 
including a factor of $(-i)$.
It is thus evident that the propagator is given by 
$P(p)^\mu{}_\nu=-iS^{-1}(p)^\mu{}_\nu$. 
With Eq.~\rf{inverse}, 
we obtain an exact, explicit expression for the modified
Stueckelberg propagator in momentum space:
\begin{multline}
\label{propagator}
{}\hspace{-3mm}P=\frac{-i}{\det(S)}\left(
\tfrac{1}{3}[S^3]\openone+\tfrac{1}{6}[S]^3\openone-\tfrac{1}{2}[S^2][S]\openone\right.\\
\left.{}-\tfrac{1}{2}[S]^2S+\tfrac{1}{2}[S^2]S
+[S]\, S^2
-S^3\right).
\end{multline}
As before, 
$S^n$ denotes the $n$th matrix power of $S(p)^\mu{}_\nu$,
and the matrix trace is abbreviated by square brackets. 
We mention that 
restricting each of the infinite sums in Eq.~\rf{smecoeff} to their first term
and setting both $m$ and $\hat{G}$ to zero 
yields the limit 
in which previous propagator expressions
have been considered~\cite{prop}.

When the exact tree-level propagator~\rf{propagator} 
is Fourier-transformed to position space, 
an integration contour must be selected.
As in the conventional case, 
this choice depends on the boundary conditions
(e.g., retarded, advanced, Feynman, etc.).
In the present case, 
a further issue arises.
As per our earlier assumption in Sec.~\ref{fullSME},
the Lorentz-violating terms are to be treated 
as perturbations of the usual Lorentz-invariant solutions, 
and we already commented on 
the need to eliminate spurious modes.
In the present context concerning the propagator, 
this may, 
for example, 
be achieved with a careful choice of (counter)clockwise integration contours that
encircle only the desired poles for each of the two orientations.
We remark that 
these issues are absent when 
the model is restricted to terms of mass dimension three and four.


\section{Leading-order results}
\label{rules} 

The absence of compelling observational evidence for 
departures from Lorentz and CPT symmetry in nature implies that
if the coefficients $\hat{k}_{F}$ and $\hat{k}_{AF}$ are nonzero,
they must be extremely small. 
The same reasoning holds true for $\hat{G}$, 
at least in the context of the $Z^0$ and $W^\pm$ bosons~\cite{fn5}. 
This fact is consistent with the theoretical expectation that
potential Lorentz and CPT violation in nature 
would be heavily suppressed,
for example by at least one power of the Planck scale.
For most phenomenological studies and many theoretical investigations 
it is therefore justified to drop higher orders in $\hat{k}_{F}$, $\hat{k}_{AF}$, and $\hat{G}$.
This section contains a brief discussion 
of a few leading-order results. 

We begin by considering the physical piece $Q$ of the dispersion relation 
given by Eq.~\rf{physDR}. 
We remind the reader that 
$Q=0$ cannot be considered a true cubic equation in the variable $(p^2-m^2)$ 
due to the $p$ dependence of the coefficients $r_0$, $r_1$, and $r_2$.
We can nevertheless employ the expressions $R_j$, 
$j\in\{1,2,3\}$, 
for the roots of a cubic 
to transform the single equation $Q=0$
into an equivalent set of three equations:
\beq{DRexpression2}
p^2-m^2=R_j(\omega^{a,\pm}_{\vec{p}},\vec{p},\hat{k}_{AF},\hat{k}_{F},\hat{G})\,.
\eeq
Just as for the original expression $Q=0$,
it seems unfeasible to determine the {\em exact}
dispersion-relation roots from the above three equations:
the plane-wave frequency $\omega^{a,\pm}_{\vec{p}}$
contained in the 4-momentum $p^{\mu}=(\omega^{a,\pm}_{\vec{p}},\vec{p})$ 
still appears on both sides of Eq.~\rf{DRexpression2}.
In particular, 
this set of equations~\rf{DRexpression2} 
will in general possess more than six solutions, 
which is consistent with the fact that 
$Q=0$ fails to be a true cubic
and can still contain a multitude of spurious Ostrogradski modes.

Before continuing, 
one may ask whether the six physical solutions
can exhibit undesirable features.
A detailed analysis of this question would be interesting 
but lies outside our present scope.
However, 
continuity implies that 
small Lorentz violation leads to small deviations from the conventional 
dispersion-relation branches,
so that the solutions must in general be well behaved 
within the validity range of the our effective field theory. 
Exceptions from this expectation 
require non-generic circumstances, 
such as particular parameter combinations.
For example, 
in the minimal SME 
only a single coefficient---the timelike $k^{(3)}_{AF}$---is known
to be problematic under some circumstances~\cite{SMEpapers}.
At the level of the dispersion relation,
this is reflected in the presence of complex-valued solutions. 
For example,
if $m=0$ 
and $k^{(3)}_{AF}$ is the only non-zero Lorentz-violating coefficient,
we have
\beq{ImExample}
R_2=-2(k^{(3)}_{AF})^2-2\sqrt{(k^{(3)}_{AF})^4+(p\cdot k^{(3)}_{AF})^2}\,.
\eeq
When $k^{(3)}_{AF}$ is timelike, 
the $j=2$ contribution in Eqs.~\rf{DRexpression2} 
cannot have real solutions $\omega_{\vec{p}}$ for small enough $\vec{p}$:
Suppose there were real solutions $\omega_{\vec{p}}\in \mathbb{R}$, 
then the square root would be real, 
and $R_2$ would be real and negative.
For the special value $\vec{p}=\vec{0}$, 
this would give $\omega_{\vec{0}}{}^2=R_2(\vec{p}=\vec{0})<0$, 
which contradicts our assumption $\omega_{\vec{p}}\in \mathbb{R}$. 
In what follows, 
we disregard such isolated regions in SME parameter space. 

The physical solutions have to represent small perturbations to 
the Lorentz-symmetric dispersion relation $p^2-m^2=0$.
Equation~\rf{DRexpression2} establishes that 
the departures from the conventional case  
are controlled by the three $R_j$. 
In a concordant frame~\cite{caus}, 
this means that
(as opposed to the spurious Ostrogradski modes)
the physical roots are characterized by 
small $R_j$ that
approach zero in the limit of vanishing Lorentz breaking.
This observation suggests solving Eq.~\rf{DRexpression2} perturbatively by 
introducing a parameter $\lambda$
multiplying the $R_j$, 
and writing a Taylor expansion for the plane-wave frequencies:
\beq{perturbative-Taylor-series}
\omega^{j,\pm}_{\vec{p}}=\omega^{j,\pm}_{\vec{p},0} + \lambda\omega^{j,\pm}_{\vec{p},1}
+ \lambda^2\omega^{j,\pm}_{\vec{p},2}+\ldots\,,
\eeq
where we have identified the label $j$ of the root expression 
with the label $a$ of the plane-wave frequency.
Substituting~\rf{perturbative-Taylor-series} on both sides of Eq.~\rf{DRexpression2} 
and expanding in powers of $\lambda$
yields an expression 
from which the coefficients of the Taylor series~\rf{perturbative-Taylor-series} 
can be determined by matching the appropriate terms. 
Solutions to Eq.~\rf{DRexpression2} are then obtained by taking $\lambda=1$, 
and the Lorentz-invariant roots $\omega^{j,\pm}_{\vec{p}}=\omega^{j,\pm}_{\vec{p},0}$ 
are recovered for $\lambda=0$.
Thus, 
this procedure continuously connects six exact dispersion-relation solutions---two for each $j$ 
in Eq.~\rf{DRexpression2}---to the Lorentz-symmetric mass shell.
This behavior is consistent with that of the physical roots, 
so that the procedure simultaneously serves as 
a filter for rejecting the spurious modes, 
which cannot remain finite in the $\lambda \to 0$ limit.

A useful practical way to solve for the coefficients in Expansion~\rf{perturbative-Taylor-series}
is iteration: 
substitute as a first iteration
$\omega^{j,\pm}_{\vec{p}}=\omega^{j,\pm}_{\vec{p},0}$ on the right-hand side of
Eq.~\rf{DRexpression2},
which determines the improved value $\omega^{j,\pm}_{\vec{p},1}$ for 
the plane-wave frequency 
on the left-hand side of this equation.
Repeating this process yields
\beq{iteration}
\omega^{j,\pm}_{\vec{p}, n+1}=
\pm\sqrt{\vec{p}^{\,2}+m^2
+R_j(\omega^{j,\pm}_{\vec{p}, n},\vec{p},\hat{k}_{AF},\hat{k}_{F},\hat{G})}\;,
\eeq
where we have denoted the iterative step by the subscript $n$
on $\omega^{j,\pm}_{\vec{p},n}$. 
One can check that 
the $n$th iterative step possesses the correct value for
(at least) 
the first $n$ coefficients in Eq.~\rf{perturbative-Taylor-series}.
In this process, 
there is actually no need to introduce explicitly the
parameter $\lambda$, 
as we can take it equal to unity immediately.

Note that 
particular care may be required for special regions in momentum space.
One example concerns very small $m$ and $\vec{p}$. 
The exact branches associated to the conventional 
positive- and negative-valued mass shell 
could then lie closely together, 
so that convergence to the correct branch 
must be ensured. 
Moreover, 
the exact solution for a positive branch 
may briefly dip into the negative spectrum 
and vice versa
leading to an unequal number of positive and negative roots 
in this small momentum-space region. 
The signs of the square root in the recursive equation~\rf{iteration} 
may then have to be adjusted correspondingly. 
Other issues, 
such as isolated momentum-space points 
at which the $\lambda$ expansion of $s_1$ in the Eq.~\rf{roots} 
starts at ${\cal O}(\lambda)$, 
may be resolved by employing a different perturbation scheme.

Since $j\in\{1,2,3\}$ 
and since for large enough $\vec{p}$ 
there is a positive and a negative square root for each $j$, 
there are six
equations 
corresponding to six seemingly independent physical plane-wave frequencies for a given $\vec{p}$.
However, 
the behavior of the modified Stueckelberg operator $S(p)$
under the replacement $p\to-p$
leads to a correspondence between the positive- and negative-root solutions.
This implies that 
there are in fact only three independent polarization modes,
as expected for a massive spin-1 field. 
More details regarding this line of reasoning 
can be found in Appendix~\ref{symmetries}.

Next, 
we consider the momentum-space propagator given by Eq.~\rf{propagator}. 
This propagator describes two physical features. 
First, 
the poles of the denominator select the plane-wave momenta.
In a Feynman-diagram context, 
for instance,
Lorentz-violating corrections to the usual poles 
can cause the propagator to go on-shell,
possibly allowing processes that
are forbidden in the Lorentz-symmetric case.
Second, 
the numerator governs features associated with the polarization states. 
For example, 
Lorentz-violating numerator corrections could yield suppressed processes that 
would be forbidden by angular-momentum conservation 
in the corresponding Lorentz-invariant situation.

If corrections to both of these features are to be retained,
the denominator and the numerator in Eq.~\rf{propagator} 
should be expanded separately to leading order. 
With 
such an approximation,  
the dominant contribution to 
the propagator 
can be separated into the 
two pieces: 
\begin{widetext}
\bea{ApprProp}
P^{\mu\nu} & = & -i(p^2-m^2)\,
\frac{\big(p^2-m^2+2[\hat{K}]
+\hat{G}^{p}_{p}-m^2 [\hat{G}]\big)\eta^{\mu\nu}
-2\hat{K}^{\mu\nu}
+2i\hat{\cal E}^{\mu\nu}
+m^2\hat{G}^{\mu\nu}
+p^\mu\hat{G}^\nu_p+p^\nu\hat{G}^\mu_p}{\prod_{j=1}^3\big[p^2-m^2
-R_j(\sqrt{\vec{p}^{\,2}+m^2},\vec{p},\hat{k}_{AF},\hat{k}_{F},\hat{G})\big]}\nonumber\\
&&{}+i(1-\xi)\,\frac{(p^2-m^2)\big(p^2-m^2+2[\hat{K}]-m^2 [\hat{G}]\big)}
{(\hat{\eta}^{\mu\nu}p_\mu p_\nu-\xi m^2)\prod_{j=1}^3\big[p^2-m^2
-R_j(\sqrt{\vec{p}^{\,2}+m^2},\vec{p},\hat{k}_{AF},\hat{k}_{F},\hat{G})\big]}\;p^\mu p^\nu
+\textrm{higher order.}
\eea
\end{widetext}
In the denominators, 
we have used the leading-order expressions for the $R_j$
implementing our previous results 
for the dispersion relation.

We note that 
the first piece of the propagator expression 
is independent of $\xi$ 
and only exhibits poles corresponding to the three physical modes.
The second piece is $\xi$ dependent 
and contains the additional pole 
associated to the auxiliary mode.
As expected, 
the second piece is without physical effects:
contraction with a source $\bar{\jmath}^{\,\nu}$ 
yields an overall factor $p\cdot \bar{\jmath}$, 
which vanishes for conserved currents.
In any case,
all contributions from the second piece 
would necessarily have to be proportional to $(p\cdot \bar{\jmath}\,)\,p^\mu$, 
which is pure gauge.
Moreover, 
the second term vanishes identically 
in Feynman--'t Hooft gauge $\xi=1$.
We further remark that 
the expression~\rf{ApprProp} does not propagate 
the additional multitude of unphysical higher-derivative modes 
discussed in the previous section.
This is required for a smooth behavior 
in the limit of vanishing Lorentz breakdown. 
Some remarks regarding the massless limit of the propagator~\rf{ApprProp} 
are contained in Appendix~\ref{massless}.

In many situations, 
Lorentz-violating corrections to the pole structure of the propagator 
can be disregarded. 
Then, 
the expression for propagator can be simplified even further with
an approximation that 
essentially corresponds to treating the Lorentz-violating pieces in the Lagrangian
as interaction terms. 
This alternative approximation is determined next.

\begin{figure}[t]
\begin{center}
\vskip+10pt
\includegraphics[width=0.7\hsize]{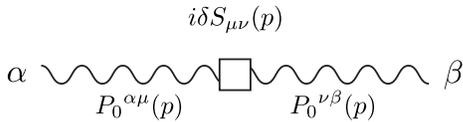}
\end{center}
\caption{Lorentz-violating propagator insertion.
The wavy lines represent the usual Lorentz-symmetric Stueckelberg propagator 
given by Eq.~\rf{P0}. 
The square box denotes the the Lorentz-breaking insertion 
determined by Eq.~\rf{delta-S}.
\vspace{-5mm}
}
\label{insertion}
\end{figure}

We begin by  considering the Lorentz-invariant part $S_0$ of Stueckelberg operator
\begin{equation}
\label{S0}
S_0(p)^{\mu\nu}=(p^2 - m^2)\,\eta^{\mu\nu}
- \left(1 - {1 \over \xi} \right) p^\mu p^\nu\,.
\end{equation}
The expression for the  corresponding momentum-space propagator 
is given by~\cite{IZ}:
\begin{equation}
\label{P0}
P_0(p)^{\mu\nu} =-i\,\frac{\eta^{\mu\nu}-p^\mu p^\nu/m^2}{p^2- m^2 }
- i\,\frac{p^\mu p^\nu/m^2}{p^2 - \xi m^2 }\,.
\end{equation}
Starting from these zeroth-order expressions,
we may now decompose the full Stueckelberg operator as
\begin{equation}
S(p)^{\mu\nu}=S_0(p)^{\mu\nu}-\delta S(p)^{\mu\nu}\,,
\end{equation}
with
\bea{delta-S}
{}\hspace{-5mm}
\delta S(p)^{\mu\nu} & = &
2i\hat{\cal E}^{\mu\nu}-2\hat{K}^{\mu\nu}+m^2\hat{G}^{\mu\nu}\nonumber\\
&& {}-\frac{1}{\xi}\big(\hat{G}^\mu_p\, p^{\nu}+\hat{G}^\nu_p\, p^{\mu}+\hat{G}^\mu_p\hat{G}^\nu_p\big)\,.
\eea
Suppressing Lorentz indices and the dependence on $p$ for brevity, 
we can now expand the full propagator $P$ 
about its Lorentz-symmetric value $P_0$ as follows:
\bea{perturbative-P}
P
&=& -i(S_0 - \delta S)^{-1}\nonumber \\
&=&-i\left[  \left(1 - \delta S\; S_0^{-1} \right) S_0\right]^{-1}\nonumber \\
&=&-iS_0^{-1} \left( 1 - \delta S\; S_0^{-1} \right)^{-1}\nonumber \\
&=& P_0 \biggl( 1 +\sum_{n=1}^\infty (i\delta S\; P_0)^n\biggr)\nonumber \\
&=& P_0 + P_0\;i\delta S\;P_0 + P_0\;i\delta S\;P_0\;i\delta S\;P_0+ \ldots\,,\qquad
\eea
where we used $P_0=-iS_0^{-1}$.
The infinite geometric series should be convergent
for small enough Lorentz violation $\delta S$.
In the context of quantum-field perturbation theory, 
the expansion~\rf{perturbative-P}
can be represented diagrammatically 
with a propagator insertion $i\delta S$ 
into internal photon lines, 
as shown in Fig.~\ref{insertion}. 
This result appears to be consistent with Matthews' theorem~\cite{matthews}.
Like the previous expression~\rf{ApprProp},
the expansion~\rf{perturbative-P} of the propagator 
around its Lorentz-symmetric value
automatically accomplishes the elimination of the unphysical poles.

\section{Summary}
\label{summary}

In this work,
we have employed the Stueckelberg method 
to introduce a mass term 
into the full SME's photon sector 
in an $R_\xi$-type gauge, 
such that the usual smooth behavior of the model 
in the $m\to 0$ limit is maintained. 
We have discussed the resulting equations of motion 
in the presence of external conserved sources, 
and studied their solutions.
Our results include the exact dispersion relation and propagator 
for free massive photons
incorporating all possible Lorentz-violating, 
translation- and gauge-invariant Lagrangian contributions
at arbitrary mass dimension.
From these exact results, 
we have extracted leading-order expressions 
for the dispersion-relation roots and the propagator. 
The method for obtaining the roots is recursive;
it also yields higher-order corrections, 
and it naturally separates the 
physical solutions from
spurious modes arising from 
higher-derivative terms in the Lagrangian.
Finally, 
we have verified that
the number of physical modes is three,
as expected for a massive vector particle.

Our results are intended primarily 
to provide a flexible tool for 
regularizing infrared divergences 
in quantum electrodynamics 
in the presence of general Lorentz violation.
A follow-up work in this context 
is already in preparation~\cite{CLP}.
In addition,
our expressions for the dispersion relation and propagator 
also hold for the quadratic part of the SME's 
heavy-boson sector, 
and are therefore expected to find applications 
in theoretical and phenomenological studies 
of electroweak physics with Lorentz breaking.
In the $m\to 0$ limit, 
our results yield the previously unknown massless photon propagator
in the full SME, 
a quantity indispensable for 
many Lorentz-symmetry investigations 
in electrodynamics.


\acknowledgments 
This work has been supported in part 
by the Portuguese Funda\c c\~ao para a Ci\^encia e a Tecnologia, 
by the Mexican RedFAE,
by Universidad Andr\'es Bello under Grant No.~UNAB DI-27-11/R, 
as well as by the Indiana University Center for Spacetime Symmetries.


\appendix

\section{Matrix relations}
\label{matrix}

This appendix provides a collection of various textbook linear-algebra results 
necessary for the line of reasoning in the main text. 
For completeness, 
we have sketched the derivation of these results.

Consider an arbitrary $4\times 4$ matrix $\Lambda$
and denote its four (not necessarily distinct) eigenvalues by $\lambda_j$, where $j=1,\ldots,4$.
The characteristic polynomial $C(\lambda)$ of $\Lambda$ 
can then be expressed as 
\beq{ChPoly}
C(\lambda)=(\lambda-\lambda_1)(\lambda-\lambda_2)(\lambda-\lambda_3)(\lambda-\lambda_4)\,.
\eeq
By the Cayley--Hamilton theorem, 
the matrix  $\Lambda$ must satisfy $C(\Lambda)=0$.
Expanding the right-hand side of Eq.~\rf{ChPoly} 
and replacing $\lambda\to\Lambda$ then yields
\beq{ChPoly2}
\Lambda^4+c_3\Lambda^3+c_2\Lambda^2+c_1\Lambda+c_0 \openone =0\,,
\eeq
where the coefficients $c_j$ are given by
\bea{coefficients}
{}\hspace{-5mm}c_0 \!& = &\! \lambda_1\lambda_2\lambda_3\lambda_4\,,\nonumber\\
{}\hspace{-5mm}c_1 \!& = &\! -\lambda _1 \lambda _2 \lambda _3-\lambda _1 \lambda _4 \lambda _3-
\lambda _2 \lambda _4 \lambda _3-\lambda _1 \lambda _2 \lambda _4\,,\nonumber\\
{}\hspace{-5mm}c_2 \!& = &\! \lambda _1 \lambda _2+\lambda _3 \lambda _2+\lambda _4 \lambda _2
+\lambda _1 \lambda _3+\lambda _1 \lambda _4+\lambda _3\lambda _4\,,\nonumber\\
{}\hspace{-5mm}c_3  \!& = &\! -\lambda _1-\lambda _2-\lambda _3-\lambda _4\,.
\eea

The next task is to express the $c_j$ 
in terms of manifestly component-independent expressions involving $\Lambda$. 
To this end, 
we transform $\Lambda$ to its Jordan normal form $\Lambda'=M^{-1}\Lambda M$, 
where $M$ is an appropriate $4\times4$ matrix.
Since such a similarity transformation leaves unchanged the eigenvalues, 
it is straightforward to verify that
\bea{DetTrace}
\det(\Lambda') \!& = &\! \lambda_1\lambda_2\lambda_3\lambda_4\,,\nonumber\\
\textrm{Tr}(\Lambda'^n) \!& = &\! \lambda _1^n+\lambda _2^n+\lambda _3^n+\lambda _4^n\,,
\eea
where $n\in\mathbb{N}$.
Inspection now establishes that
\bea{c}
c_0 \!& = &\! \det(\Lambda')\,,\nonumber\\
c_1 \!& = &\! \tfrac{1}{2}\textrm{Tr}(\Lambda'^2)\textrm{Tr}(\Lambda')
-\tfrac{1}{3}\textrm{Tr}(\Lambda'^3)-\tfrac{1}{6}\textrm{Tr}(\Lambda')^3\,,\nonumber\\
c_2 \!& = &\! \half\left[\textrm{Tr}(\Lambda')^2-\textrm{Tr}(\Lambda'^2)\right]\,,\nonumber\\
c_3 \!& = &\! -\textrm{Tr}(\Lambda')\,.
\eea
Since both the determinant as well as the trace of a square matrix
remain unchanged under similarity transformations, 
the relations in Eqs.~\rf{DetTrace} and~\rf{c} hold, in fact, 
also for our original $4\times 4$ matrix $\Lambda$.
Employing these results in 
Eq.~\rf{ChPoly2} gives 
\begin{multline}
\label{MatrixRelation}
\Lambda^4-[\Lambda]\, \Lambda^3
+\tfrac{1}{2}\left([\Lambda]^2-[\Lambda^2]\right)\Lambda^2\\
{}+\left(
\tfrac{1}{2}[\Lambda^2][\Lambda]-\tfrac{1}{3}[\Lambda^3]-\tfrac{1}{6}[\Lambda]^3
\right)\Lambda+ \det(\Lambda) \openone=0\,.
\end{multline}
Here, 
we have abbreviated the matrix trace by square brackets.
This result establishes that
the set of matrices $\{\openone,\Lambda,\Lambda^2,\Lambda^3,\Lambda^4\}$
is linearly dependent.

Equation~\rf{MatrixRelation} can be employed 
to determine further relations involving the $4\times 4$ matrix $\Lambda$.
For example, 
taking the trace of Eq.~\rf{MatrixRelation} yields the expression 
\bea{detexpr}
\det(\Lambda) \!& = &\! \tfrac{1}{24}[\Lambda]^4
-\tfrac{1}{4}[\Lambda^4]
+\tfrac{1}{8}[\Lambda^2]^2\nonumber\\
&&\!{}+\tfrac{1}{3}[\Lambda][\Lambda^3]
-\tfrac{1}{4}[\Lambda]^2[\Lambda^2]
\eea
for the determinant of $\Lambda$.
Another relation can be obtained by multiplying Eq.~\rf{MatrixRelation}
with the inverse matrix $\Lambda^{-1}$.
The resulting equation can then be solved for $\Lambda^{-1}$:
\begin{multline}
\label{inverse}
{}\hspace{-3mm}\Lambda^{-1}=\frac{1}{\det(\Lambda)}\left(
\tfrac{1}{3}[\Lambda^3]\openone+\tfrac{1}{6}[\Lambda]^3\openone-\tfrac{1}{2}[\Lambda^2][\Lambda]\openone\right.\\
\left.{}-\tfrac{1}{2}[\Lambda]^2\Lambda+\tfrac{1}{2}[\Lambda^2]\Lambda
+[\Lambda]\, \Lambda^2
-\Lambda^3\right).
\end{multline}
In Eqs.~\rf{detexpr} and~\rf{inverse}
we have again abbreviated the matrix trace 
by square brackets.


\section{Dispersion relation}
\label{DR}

To present $r_0$, $r_1$, and $r_2$ 
appearing in the physical dispersion relation~\rf{physDR}, 
we introduce the following more efficient notation.
We suppress the $AF$ subscript on $\hat{k}_{AF}$ 
and drop the caret:
\beq{KEDef}
(\hat{k}_{AF})^\mu\equiv k^\mu\,.
\eeq
It is also convenient to extract the traceless part $\hat{\underline{G}}{}^{\mu\nu}$ 
from the full $\hat{G}^{\mu\nu}$:
\beq{TracelessDef}
\hat{\underline{G}}{}^{\mu\nu}\equiv\hat{G}^{\mu\nu}-\tfrac{1}{4}[\hat{G}]\eta^{\mu\nu}\,.
\eeq
Here, 
we have again denoted the matrix trace by 
square brackets 
$[\hat{G}]\equiv\hat{G}^{\mu}_{\mu}$.
Inspection of our model Lagrangian~\rf{eq:totalL} reveals that 
for the photon the decomposition~\rf{TracelessDef} 
can be implemented by the replacements
\bea{replace}
\hat{G}^{\mu\nu} & \to & \underline{\hat{G}}{}^{\mu\nu}\nonumber\\
m & \to & \hat{m} \equiv m\big(1+\tfrac{1}{4}[\hat{G}]\big)\nonumber\\
\xi^{-1} & \to & \hat{\xi}^{-1}\equiv\xi^{-1}\big(1+\tfrac{1}{4}[\hat{G}]\big)^2\,,
\eea
where it is understood that
$\hat{\xi}^{-1/2}$ has to be placed at the appropriate position 
in the gauge-fixing term.
We also remind the reader that
we denote the contraction of a 4-vector with a symmetric 2-tensor 
by writing the vector as a superscript or a subscript on the tensor:
for example $(\underline{\hat{G}}{}^2){}^k_p\equiv 
\underline{\hat{G}}{}^\alpha_\beta 
\underline{\hat{G}}{}^\beta_\gamma\,  p_\alpha (\hat{k}_{AF})^\gamma$, etc.

With this notation, 
the $r$ coefficients in the physical dispersion relation~\rf{physDR} 
take the following form:
\begin{widetext}
\bea{physCoeff}
{}\hspace{-10mm}
r_0 \!& = &\!
\tfrac{4}{3}[\hat{K}]^3+\tfrac{8}{3}[\hat{K}^3]-4[\hat{K}][\hat{K}^2]+8\hat{m}^2\hat{K}^k_k+2[\hat{K}]^2\underline{\hat{G}}{}^p_p
-2[\hat{K}^2]\underline{\hat{G}}{}^p_p-4(k\cdot p)^2 \underline{\hat{G}}{}^p_p-4\hat{m}^2[\underline{\hat{G}}{}\hat{K}^2]+4\hat{m}^2[\hat{K}][\underline{\hat{G}}{}\hat{K}]
\nonumber\\
&&\!{}
-2\hat{m}^2(\underline{\hat{G}}{}\hat{K}\underline{\hat{G}}{})^p_p+2\hat{m}^2[\hat{K}](\underline{\hat{G}}{}^2)^p_p+2\hat{m}^2[\underline{\hat{G}}{}\hat{K}]\underline{\hat{G}}{}^p_p
+8\hat{m}^2(k\cdot p)\underline{\hat{G}}{}^k_p+2\hat{m}^4[\underline{\hat{G}}{}^2\hat{K}]-\hat{m}^4[\hat{K}][\underline{\hat{G}}{}^2]-4\hat{m}^4\underline{\hat{G}}{}^k_k
\nonumber\\
&&\!{}
+\hat{m}^4(\underline{\hat{G}}{}^3)^p_p-\tfrac{1}{2}\hat{m}^4[\underline{\hat{G}}{}^2]\underline{\hat{G}}{}^p_p
-\tfrac{1}{3}\hat{m}^6[\underline{\hat{G}}{}^3]-8[\underline{\hat{G}}{}\hat{K}^3]+8[\hat{K}][\underline{\hat{G}}{}\hat{K}^2]-4[\hat{K}]^2[\underline{\hat{G}}{}\hat{K}]+4[\hat{K}^2][\underline{\hat{G}}{}\hat{K}]+8\hat{K}^k_k \underline{\hat{G}}{}^p_p
\nonumber\\
&&\!{}
+4\hat{m}^2[\underline{\hat{G}}{}^2\hat{K}^2]-2\hat{m}^2[\underline{\hat{G}}{}\hat{K}]^2+2\hat{m}^2[(\underline{\hat{G}}{}\hat{K})^2]
+\hat{m}^2[\hat{K}]^2[\underline{\hat{G}}{}^2]-\hat{m}^2[\hat{K}^2][\underline{\hat{G}}{}^2]-4\hat{m}^2[\hat{K}][\underline{\hat{G}}{}^2\hat{K}]+4\hat{m}^2(\underline{\hat{G}}{}^k_p)^2
\nonumber\\
&&\!{}
-4\hat{m}^2\underline{\hat{G}}{}^k_k\underline{\hat{G}}{}^p_p-2\hat{m}^4[\underline{\hat{G}}{}^3\hat{K}]+\hat{m}^4[\underline{\hat{G}}{}\hat{K}][\underline{\hat{G}}{}^2]
+\tfrac{2}{3}\hat{m}^4[\hat{K}][\underline{\hat{G}}{}^3]-\tfrac{1}{8}\hat{m}^6[\underline{\hat{G}}{}^2]^2+\tfrac{1}{4}\hat{m}^6[\underline{\hat{G}}{}^4]\,,
\nonumber\\
{}\hspace{-10mm}r_1 \!& = &\!
2[\hat{K}]^2-2[\hat{K}^2]-4(k\cdot p)^2+4\hat{m}^2k^2+2[\hat{K}]\underline{\hat{G}}{}^p_p+2\hat{m}^2[\underline{\hat{G}}{}\hat{K}]+\hat{m}^2(\underline{\hat{G}}{}^2)^p_p
-\tfrac{1}{2}\hat{m}^4[\underline{\hat{G}}{}^2]+8\hat{K}^k_k+4k^2\underline{\hat{G}}{}^p_p-4\hat{m}^2\underline{\hat{G}}{}^k_k\,,
\nonumber\\
{}\hspace{-10mm}r_2 \!& = &\!
2[\hat{K}]+\underline{\hat{G}}{}^p_p+4k^2
\,.
\eea
The above dispersion relation can be shown to reduce in the appropriate limit
to mass-dimension three results quoted in Ref.~\cite{AACGS}.

To determine the quantities $R_j$ appearing in Eq.~\rf{DRexpression2}, 
we apply the usual Tschirnhaus-type transformation for cubic equations 
to Expression~\rf{physDR}. 
This yields the equivalent depressed form 
of the physical dispersion relation $Q=0$:
\beq{Tschirnhaus}
\left(p^2-\hat{m}^2+\tfrac{2}{3}[\hat{K}]+\tfrac{1}{3}\underline{\hat{G}}{}^p_p+\tfrac{4}{3}k^2\right)^3
+s_1\left(p^2-\hat{m}^2+\tfrac{2}{3}[\hat{K}]+\tfrac{1}{3}\underline{\hat{G}}{}^p_p+\tfrac{4}{3}k^2\right)
+s_0=0\,,
\eeq
where coefficients $s_0$ and $s_1$ are given by
\bea{sCoeff}
{}\hspace{-10mm}
s_0 
\!& = &\!
\tfrac{16}{27}[\hat{K}]^3+\tfrac{8}{3}[\hat{K}^3]-\tfrac{8}{3}[\hat{K}][\hat{K}^2]+\tfrac{8}{3}(k\cdot p)^2[\hat{K}]
+8\hat{m}^2\hat{K}^k_k-\tfrac{8}{3}\hat{m}^2k^2 [\hat{K}]+\tfrac{8}{9}[\hat{K}]^2\underline{\hat{G}}{}^p_p
-\tfrac{4}{3}[\hat{K}^2]\underline{\hat{G}}{}^p_p-\tfrac{2}{9}[\hat{K}](\underline{\hat{G}}{}^p_p)^2\qquad
\nonumber\\
&&\!{}
-\tfrac{8}{3}(k\cdot p)^2 \underline{\hat{G}}{}^p_p+\tfrac{2}{27}(\underline{\hat{G}}{}^p_p)^3-4\hat{m}^2[\underline{\hat{G}}{}\hat{K}^2]
+\tfrac{8}{3}\hat{m}^2[\hat{K}][\underline{\hat{G}}{}\hat{K}]-2\hat{m}^2(\underline{\hat{G}}{}\hat{K}\underline{\hat{G}}{})^p_p+\tfrac{4}{3}\hat{m}^2[\hat{K}](\underline{\hat{G}}{}^2)^p_p
+\tfrac{4}{3}\hat{m}^2[\underline{\hat{G}}{}\hat{K}]\underline{\hat{G}}{}^p_p
\nonumber\\
&&\!{}
-\tfrac{1}{3}\hat{m}^2\underline{\hat{G}}{}^p_p(\underline{\hat{G}}{}^2)^p_p-\tfrac{4}{3} \hat{m}^2 k^2 \underline{\hat{G}}{}^p_p
+8\hat{m}^2(k\cdot p)\underline{\hat{G}}{}^k_p+2\hat{m}^4[\underline{\hat{G}}{}^2\hat{K}]-\tfrac{2}{3}\hat{m}^4[\hat{K}][\underline{\hat{G}}{}^2]
-4\hat{m}^4\underline{\hat{G}}{}^k_k-\tfrac{1}{3}\hat{m}^4[\underline{\hat{G}}{}^2]\underline{\hat{G}}{}^p_p
\nonumber\\
&&\!{}
+\hat{m}^4(\underline{\hat{G}}{}^3)^p_p-\tfrac{1}{3}\hat{m}^6[\underline{\hat{G}}{}^3]+\tfrac{8}{9}k^2[\hat{K}]^2
+\tfrac{8}{3}k^2[\hat{K}^2]-\tfrac{16}{3}[\hat{K}]\hat{K}^k_k+\tfrac{16}{3}k^2(k\cdot p)^2
-\tfrac{4}{9}k^2 (\underline{\hat{G}}{}^p_p)^2+8[\hat{K}][\underline{\hat{G}}{}\hat{K}^2]
\nonumber\\
&&\!{}
-8[\underline{\hat{G}}{}\hat{K}^3]-4[\hat{K}]^2[\underline{\hat{G}}{}\hat{K}]+4[\hat{K}^2][\underline{\hat{G}}{}\hat{K}]+\tfrac{16}{3}\hat{K}^k_k \underline{\hat{G}}{}^p_p
-\tfrac{16}{9}k^2 [\hat{K}] \underline{\hat{G}}{}^p_p-4\hat{m}^2[\hat{K}][\underline{\hat{G}}{}^2\hat{K}]+2\hat{m}^2[(\underline{\hat{G}}{}\hat{K})^2]
\nonumber\\
&&\!{}
+4\hat{m}^2[\underline{\hat{G}}{}^2\hat{K}^2]-2\hat{m}^2[\underline{\hat{G}}{}\hat{K}]^2+\hat{m}^2[\hat{K}]^2[\underline{\hat{G}}{}^2]-\hat{m}^2[\hat{K}^2][\underline{\hat{G}}{}^2]
-\tfrac{8}{3}\hat{m}^2 k^2 [\underline{\hat{G}}{}\hat{K}]+4\hat{m}^2(\underline{\hat{G}}{}^k_p)^2-\tfrac{8}{3}\hat{m}^2\underline{\hat{G}}{}^k_k\underline{\hat{G}}{}^p_p
\nonumber\\
&&\!{}
-\tfrac{16}{3}\hat{m}^2 k^4+\tfrac{8}{3}\hat{m}^2[\hat{K}]\underline{\hat{G}}{}^k_k-\tfrac{4}{3}\hat{m}^2 k^2 (\underline{\hat{G}}{}^2)^p_p
-2\hat{m}^4[\underline{\hat{G}}{}^3\hat{K}]+\hat{m}^4[\underline{\hat{G}}{}\hat{K}][\underline{\hat{G}}{}^2]+\tfrac{2}{3}\hat{m}^4[\hat{K}][\underline{\hat{G}}{}^3]
+\tfrac{2}{3}\hat{m}^4 k^2 [\underline{\hat{G}}{}^2]
\nonumber\\
&&\!{}
-\tfrac{1}{8}\hat{m}^6[\underline{\hat{G}}{}^2]^2+\tfrac{1}{4}\hat{m}^6[\underline{\hat{G}}{}^4]+\tfrac{64}{9}k^4[\hat{K}]
-\tfrac{32}{3}k^2 \hat{K}^k_k-\tfrac{16}{9}k^4 \underline{\hat{G}}{}^p_p+\tfrac{16}{3}\hat{m}^2 k^2 \underline{\hat{G}}{}^k_k
+\tfrac{128}{27}k^6\,,
\nonumber\\
{}\hspace{-10mm}s_1 
\!& = &\!
\tfrac{2}{3}[\hat{K}]^2-2[\hat{K}^2]-4(k\cdot p)^2+4\hat{m}^2k^2+\tfrac{2}{3}[\hat{K}]\underline{\hat{G}}{}^p_p-\tfrac{1}{3}(\underline{\hat{G}}{}^p_p)^2
+2\hat{m}^2[\underline{\hat{G}}{}\hat{K}]+\hat{m}^2(\underline{\hat{G}}{}^2)^p_p-\tfrac{1}{2}\hat{m}^4[\underline{\hat{G}}{}^2]+8\hat{K}^k_k
\nonumber\\
&&\!{}
-\tfrac{16}{3}k^2 [\hat{K}]+\tfrac{4}{3}k^2\underline{\hat{G}}{}^p_p-4\hat{m}^2\underline{\hat{G}}{}^k_k-\tfrac{16}{3}k^4
\,.
\eea

With $s_0$ and $s_1$ at hand, 
the usual formulae for the roots of a cubic
determine explicit expressions for the $R_j$ quantities. 
These expressions can then be used for the iteration Eq.~\rf{iteration}:
Suppose we want to calculate the first-order frequencies 
from the zeroth-order solution $(\sqrt{\vec{p}^{\,2}+m^2},\vec{p})$,
which corresponds to the first iteration step in Eq.~\rf{iteration}. 
With $(\sqrt{\vec{p}^{\,2}+m^2},\vec{p})$ and Eqs.~\rf{sCoeff}, 
we can in principle check the sign of the discriminant $4s_1{}^3+27s_0{}^2$, 
at least to leading order.
A positive discriminant 
is associated with complex-valued $R_j$, 
which would appear to lead to unphysical roots. 
We therefore take the discriminant to be non-positive 
in what follows.  
The $R_j$ can then be expressed trigonometrically:
\beq{roots}
R_j=2\sqrt{-\frac{s_1}{3}}\cos\left[
\frac{1}{3}\cos^{-1}\left(
\frac{3s_0}{2s_1}\sqrt{-\frac{3}{s_1}}
\right)
-\frac{2\pi}{3}j
\right]
-\tfrac{2}{3}[\hat{K}]-\tfrac{1}{3}\underline{\hat{G}}{}^p_p-\tfrac{4}{3}k^2\,,\quad j\in\{1,2,3\}\,.
\eeq
\end{widetext}
These explicit $R_j$  
can now,  
for example, 
be employed in the context of Eq.~\rf{iteration}. 

We remark that 
for the special case of a vanishing discriminant,
Eq.~\rf{roots} determines a multiple root, 
so that at least two physical excitations 
propagate without birefringence. 
We also note that
for non-positive discriminant 
the argument of the cosine function in Eq.~\rf{roots} 
is real, 
and its magnitude therefore remains bounded by one. 
The limit of zero Lorentz violation 
is then determined by $s_1$, $[\hat{K}]$, $\underline{\hat{G}}{}^p_p$, and $k^2$. 
These coefficients will generically be momentum dependent. 
In particular, 
the zero components for the unphysical solutions
may diverge in this limit, 
so that the corresponding behavior of the $R_j$ is unclear {\em a priori}.
However, 
for the physical solutions 
the plane-wave frequencies must remain bounded 
because they are defined to approach the conventional expressions 
in the Lorentz-symmetric limit.
For vanishing Lorentz breaking,  
the physical $R_j$ are therefore seen to approach zero,
as required by self-consistency.


\section{Discrete symmetries}
\label{symmetries}

An interesting question concerns
the number of independent plane-wave solutions.
Some insight into this issue 
can be gained by investigating discrete symmetries 
of both the plane-wave polarization vectors and 
the dispersion relation~\rf{fullDR}.

Since the Lagrangian~\rf{eq:totalL} is real,
the momentum-space Stueckelberg operator
must be hermitian
\beq{hermitian-S}
S^{\mu\nu}(p)^*=S^{\nu\mu}(p)\,,
\eeq
which can also be established directly by inspection of Eq.~\rf{Soperator}.
Next, 
we consider $S^{\mu\nu}(p)$ under $p \to -p$.
Only the $\hat{k}_{AF}$ piece reverses its sign
because it is odd in $p$.
All other terms are $p$ even 
and remain unchanged. 
We also need the behavior of $S^{\mu\nu}(p)$ under complex conjugation.
Since the position-space Stueckelberg operator is purely real,
complex-valued contributions to  
the momentum-space $S^{\mu\nu}(p)$
can only arise from the Fourier replacement $\partial\to -ip$.
We will focus on real-valued plane-wave momenta $p$
since otherwise an interpretation as a propagating solution 
is difficult. 
It then follows that 
the $p$-odd $\hat{k}_{AF}$ term is purely imaginary, 
while all other terms are even in $p$ even 
and thus real.
Complex conjugation of $S^{\mu\nu}(p)$ 
therefore only changes the sign of its $\hat{k}_{AF}$ piece.
It thus becomes evident that 
the momentum-space Stueckelberg operator is left unaffected 
under the combination of the two operations:
\bea{operations}
{}\hspace{-5mm}S^{\mu\nu}(+p,+\hat{k}_{AF},\hat{k}_{F},\hat{G}) 
&=&S^{\mu\nu}(-p,+\hat{k}_{AF},\hat{k}_{F},\hat{G})^*\nonumber\\
&=&S^{\nu\mu}(-p,+\hat{k}_{AF},\hat{k}_{F},\hat{G})\,,
\eea
where the Hermiticity of $S^{\mu\nu}$ has been used. 
Since any determinant is invariant under transposition, 
we find
\beq{DRsymm}
\det S(+p,\hat{k}_{AF},\hat{k}_{F},\hat{G})=\det S(-p,\hat{k}_{AF},\hat{k}_{F},\hat{G})\,,
\eeq
i.e., $p\to -p$ represents a discrete symmetry of the dispersion relation.
This invariance also becomes clear
when Eqs.~\rf{fullDR}, \rf{physDR}, and~\rf{physCoeff} are combined. 

Continuing our analysis in a concordant frame~\cite{caus},
we fix a wave 3-vector $\vec{p}$ 
and imagine solving the dispersion relation~\rf{fullDR}.
Following the reasoning presented in Sec.~\ref{rules}, 
we can eliminate the spurious Ostrogradski modes
and focus on the six physical branches of the dispersion-relation solution 
with exact $\omega^{j,\pm}_{\vec{p}}$. 
In these expressions, 
the $\pm$ label must correspond to the sign of the solution, 
a property that follows immediately from 
the required proximity to the Lorentz-symmetric sheets~\cite{fn3}.
Without loss of generality, 
we now consider the specific plane-wave 4-momentum 
with positive energy $(p_+^j)^\mu\equiv(\omega^{j,+}_{-\vec{p}},-\vec{p})$.
The $ p \to -p$ symmetry of the dispersion relation implies that
$(p_-^j)^\mu\equiv(-\omega^{j,+}_{-\vec{p}},\vec{p})$ 
also represents a solution, 
but one with negative frequency.
Moreover, 
this solution must be physical: 
since $p_+^j$ is close to the conventional positive sheet,
$p_-^j$ must be a perturbation of the negative sheet. 
With a suitable selection of the $j$ labels, 
we may therefore conclude $\omega^{j,-}_{\vec{p}}=-\omega^{j,+}_{-\vec{p}}$.
In summary, 
the set of physical plane-wave momenta is given by
\beq{PWMomenta}
\left\{
(p_+^j)^\mu=(\omega_{\vec{p}}^{j,+},\vec{p})\,\,,\,\,\,\,
(p_-^j)^\mu=(-\omega_{-\vec{p}}^{j,+},\vec{p})
\right\},
\eeq
and is thus determined entirely by 
the expressions for the three positive-frequency solutions $\omega^{j,+}_{\vec{p}}$.

In addition to the plane-wave momenta 
$(p_\pm^j)^\mu$, 
the associated polarization vectors 
$(\varepsilon^j_\pm)^\mu(\vec{p})$ 
are needed to construct a plane wave.
Given the  $\varepsilon^j_+(\vec{p})$ 
associated with the positive-frequency roots, 
we can construct the remaining negative-frequency $\varepsilon^j_-(\vec{p})$ 
as follows.
Up to normalization, 
the polarization vectors are by definition 
in the kernel of the Stueckelberg operator:
$S(\omega^{j,\pm}_{\vec{p}},\vec{p})\,\varepsilon^j_\pm(\vec{p})=0$. 
This definition contains 
$S(\omega^{j,+}_{-\vec{p}},-\vec{p})\,\varepsilon^j_+(-\vec{p})=0$
as a special case.
Complex conjugation combined with Eq.~\rf{operations} 
then yields $S(-\omega^{j,+}_{-\vec{p}},\vec{p})\,\varepsilon^j_+{}^*(-\vec{p})=0$.
Together with the plane-wave momentum property~\rf{PWMomenta},
this equation is identified as the defining relation for $\varepsilon^j_-(\vec{p})$.
Using suitable normalizations, 
we arrive at 
\beq{polarizations}
(\varepsilon^j_-)^\mu(\vec{p})=(\varepsilon^j_+{}^*)^\mu(-\vec{p})\,.
\eeq
We remark that
this relation continues to hold 
for dispersion-relation roots 
with higher multiplicities
when the bases of $\textrm{Ker}\,S(-\omega^{j,+}_{-\vec{p}},\vec{p})$ 
are chosen appropriately.

An arbitrary, free, real-valued, physical solution $A^\mu(x)$ 
can be represented by the following Fourier superposition 
of plane waves~\cite{fn4}: 
\bea{FourierDecomp}
\hspace{-10mm}
A^\mu(x) & = &
\int\!\!\frac{d^3\vec{p}}{(2\pi)^3}
\sum_{j=1}^{3}
\left[
a^j_+(\vec{p})\,(\varepsilon^j_+)^\mu(\vec{p})\,e^{-ip^j_+\cdot x}\right.\nonumber\\
&&\left.{}\hspace{11mm}+a^j_-(\vec{p})\,(\varepsilon^j_-)^\mu(\vec{p})\,e^{-ip^j_-\cdot x}
\right]+\textrm{c.c.}
\eea
This expression suggests that
six Fourier coefficients $\{a^j_+(\vec{p}),a^j_-(\vec{p})\}$ 
with $j=1,2,3$ must be specified independently
to determine a general solution.
However, 
a change of integration variable $\vec{p}\to-\vec{p}$ 
in the expression associated with the negative-frequency solutions 
together with Eqs.~\rf{PWMomenta} and~\rf{polarizations} yields
\beq{FourierDecomp2}
A^\mu(x) = 
\int\!\!\frac{d^3\vec{p}}{(2\pi)^3}
\sum_{j=1}^{3}
b^j(\vec{p})\,(\varepsilon^j_+)^\mu(\vec{p})\,e^{-ip^j_+\cdot x}+\textrm{c.c.},
\eeq
where we have defined $b^j(\vec{p})\equiv a^j_+(\vec{p})+a^j_-{}^*(-\vec{p})$.
This establishes that 
the six physical dispersion-relation roots 
are associated with three degrees of freedom.


\section{Massless limit}
\label{massless}

In the context of regularizing infrared divergences 
in quantum-field perturbation theory,
the massless limit of our Lorentz-violating Stueckelberg model
is of particular importance.
The photon mass was introduced into our Lagrangian 
via the Stueckelberg method,
which ensures the proper behavior in this limit. 
Nevertheless,
it is interesting to study explicitly the dispersion relation and the propagator 
of our model for vanishing $m$, 
for example, to compare with previous results for massless Lorentz-violating photons.

We begin by defining the contraction 
\beq{kafcontraction}
{}\hspace{-1.5mm}\hat{I}\equiv\epsilon_{\alpha\beta\gamma\delta}
\epsilon_{\kappa\lambda\mu\nu}
p_\rho p_\sigma p_\tau p_\iota
(\hat{k}_{F}\hspace{-.3mm}
)^{\alpha\beta\kappa\rho}
(\hat{k}_{F}\hspace{-.3mm}
)^{\lambda\sigma\tau\gamma}
(\hat{k}_{F}\hspace{-.3mm}
)^{\iota\delta\mu\nu}
\eeq
of the Lorentz-violating $\hat{k}_{F}$. 
One can show that 
the coordinate scalar $\hat{I}$ obeys the following identities:
\bea{identities}
\tfrac{1}{12}\,\hat{I}\,\underline{\hat{G}}{}^p_p & = & 2 [\underline{\hat{G}}{}\hat{K}^3]-2[\underline{\hat{G}}{}\hat{K}^2][\hat{K}]
+[\underline{\hat{G}}{}\hat{K}]\big([\hat{K}]^2-[\hat{K}^2]\big)
\nonumber\\
\tfrac{1}{4}\,\hat{I}\, p^2& = & 3 [\hat{K}][\hat{K}^2]-[\hat{K}]^3-2[\hat{K}^3]\,.
\qquad\qquad\qquad\qquad\quad
\eea
With these identities, 
the full massless dispersion relation~\rf{fullDR} can be cast into the form
\bea{masslessDR}
\hat{\xi}\det S &=& (\hat{\eta}^{\mu\nu} p_\mu p_\nu)^2
\Big(p^4+2[\hat{K}]p^2+8\hat{K}^k_k-\tfrac{1}{3}\hat{I}\nonumber\\
&&{}-4(k\cdot p)^2+4k^2 p^2+2[\hat{K}]^2-2[\hat{K}^2]
\Big)\,.\qquad
\eea
It thus becomes apparent that 
in the massless limit 
the expression $(\hat{\eta}^{\mu\nu} p_\mu p_\nu)$ can be factored out of $Q$.

The next step is to identify 
the physical part of the massless dispersion relation~\rf{masslessDR}. 
Clearly, 
one of the $(\hat{\eta}^{\mu\nu} p_\mu p_\nu)$ factors 
originates from the auxiliary mode and remains unphysical. 
To gain insight into the other factors, 
it is instructive to recall the model Lagrangian~\rf{eq:totalL}. 
In the zero-mass limit, 
$\hat{G}$ enters the equations of motion for the photon only 
via the gauge-fixing term ${\cal L}_\textrm{g.f.}$
and should therefore not lead to observable effects.
It then follows that 
the second $(\hat{\eta}^{\mu\nu} p_\mu p_\nu)$ factor 
must also be unphysical 
because it contains $\hat{\eta}=\eta+\hat{G}$.
We conclude that 
the last factor in Eq.~\rf{masslessDR}, 
which is free of $\hat{G}$,
governs the physical excitations for $m=0$. 
This factor is in agreement with 
the general photon dispersion relation 
given in Ref.~\cite{PhotonSME}, 
as required by consistency.

The exact massive propagator~\rf{propagator} 
as well as its leading-order approximation~\rf{ApprProp} 
exhibit six physical poles that 
must be encircled by the integration contours. 
The above argument identifies those two of these six poles 
that should become unphysical in the massless case.
The question then arises 
as to whether the unphysical character of these two modes
is reflected in the pole structure of the propagator 
in the $m \to 0$ limit.
This may be established by demonstrating that
the $(\hat{\eta}^{\mu\nu} p_\mu p_\nu)$ factor 
in the denominator of the propagator 
is cancelled by a corresponding $(\hat{\eta}^{\mu\nu} p_\mu p_\nu)$ factor 
in the numerator 
for vanishing mass. 
An investigation of this cancellation 
for the exact expression~\rf{propagator} 
would be interesting
but lies outside the scope of the present work. 
We instead focus on the on the leading-order piece~\rf{ApprProp} 
of the propagator, 
which is appropriate for the majority of practical applications.

In the discussion of the leading-order propagator~\rf{ApprProp},
it was already mentioned that
the $(1-\xi)$ contribution in the second line 
leaves unaffected observable quantities.
We may therefore focus on the physical contribution
contained in the first line of Eq.~\rf{ApprProp}.
In the $m \to 0$ limit, 
the numerator $-iN^{\mu\nu}$ of this first piece has the form 
\bea{numerat}
N^{\mu\nu}&=&p^2\big(2[\hat{K}]\eta^{\mu\nu}
-2\hat{K}^{\mu\nu}
+2i\hat{\cal E}^{\mu\nu}
+p^\mu\hat{G}^\nu_p+p^\nu\hat{G}^\mu_p\big)\nonumber\\
&&{}+p^2\big(p^2+\hat{G}^{p}_{p}\big)\eta^{\mu\nu}\,.
\qquad\qquad\qquad\qquad\qquad\qquad\;
\eea
The $\eta^{\mu\nu}$ term in the second line of the above Eq.~\rf{numerat} 
contains the factor $p^2+\hat{G}^{p}_{p}=\hat{\eta}^{\mu\nu} p_\mu p_\nu$. 
All terms in the first line represent leading-order corrections,
which remain unaffected by the addition of higher-order Lorentz-violating terms.
A judicious choice for such higher-order contributions is
$p^2 \to p^2+\hat{G}^{p}_{p}=\hat{\eta}^{\mu\nu} p_\mu p_\nu$ 
in the overall $p^2$ factor.
It thus becomes evident
that in the physical piece of the leading-order propagator, 
the factor $(\hat{\eta}^{\mu\nu} p_\mu p_\nu)$ 
appears in both the numerator and the denominator 
and therefore cancels. 
We conclude that
the poles that become unphysical in the massless limit
disappear from the propagator, 
at least to leading order. 

Since $\hat{G}$ arises in the massless case solely from the gauge-fixing term, 
measurable quantities are necessarily free of $\hat{G}$, 
even though $p^\mu\hat{G}^\nu_p$ and $p^\nu\hat{G}^\mu_p$ terms still appear
in the numerator of the propagator~\rf{ApprProp} and, 
equivalently, 
in the propagator insertion~\rf{delta-S}. 
When the propagator is contracted with a conserved current $\bar{\jmath}^{\,\nu}$,
$\hat{G}$ independence can be seen explicitly as follows.
For the $p^\mu\hat{G}^\nu_p$ term, 
such a contraction yields an expression proportional to $p^\mu$, 
which is pure gauge and can therefore not lead to observable effects.
For the $p^\nu\hat{G}^\mu_p$ term,
the contraction produces an overall $p\cdot\bar{\jmath}$, 
which vanishes due to current conservation. 
It is thus clear that
$\hat{G}$ is indeed unobservable in this context, 
as expected.



\begin{thebibliography}{99}
\bibitem{fn1}
Odd numbers of derivatives would be
equivalent to total-divergence terms in the Lagrangian 
piece~\rf{Stueckelberg}.
\bibitem{fn2}
This may entail a momentum cut-off, 
which is consistent with our effective-field-theory interpretation.
\bibitem{fn5}
For the photon, 
$\hat{G}$ becomes unobservable in the massless limit. 
But without loss of generality, we may take it to be small.
\bibitem{fn3}
This property may not hold for all $\vec{p}$:
in strongly boosted non-concordant coordinate systems
or for small mass-type gaps,
a branch labeled $+$ can briefly become negative
and vice versa.
However, 
this change of sign is confined to a momentum-space region 
with an extent determined by the size of Lorentz violation
and a location close to the origin. 
Such cases are already known in the minimal SME~\cite{caus}.
Although this can lead to 
unequal numbers of positive and negative roots within this small $\vec{p}$ region,
the total number of branches is left unaffected. 
This feature is irrelevant for the discussion at hand 
because it can be sidestepped by the initial selection of a large enough $\vec{p}$ 
and the extension of the final result to all $\vec{p}$ by continuity.
\bibitem{fn4}
In the conventional case,
the Fourier decomposition is often presented with 
a Lorentz-covariant integration measure 
involving the conventional plane-wave frequencies $\omega_{\vec{p}}=(\vec{p}^{\,2}+m^2)^{1/2}$. 
However, 
simply replacing $\omega_{\vec{p}} \to \omega^{j,\pm}_{\vec{p}}$ 
fails to yield a covariant expression for the measure
in the present Lorentz-violation context.
We have opted for the simplest non-covariant measure $d^3\vec{p}$
at the cost of using Fourier weights 
that do not represent coordinate scalars.
\end{thebibliography}
\end{document}